\documentclass[twocolumn,showpacs,amsmath,amssymb,floatfix,pre]{revtex4}
\usepackage{graphicx}
\begin{document}
\title{Phase transition induced for external field in tree-dimensional isotropic Heisenberg antiferromagnet} 
\author{Minos A. Neto}
\email{minos@pq.cnpq.br}
\author{J. Roberto Viana}
\author{Octavio D. R. Salmon}
\affiliation{Departamento de F\'{\i}sica, Universidade Federal do Amazonas, 3000, Japiim,
69077-000, Manaus-AM, Brazil}
\author{E. Bublitz Filho}
\affiliation{Instituto Federal de Roraima, 69303-220, Boa Vista-RR, Brazil}
\author{Jos\'{e} Ricardo de Sousa}
\affiliation{Departamento de F\'{\i}sica, Universidade Federal do Amazonas, 3000, Japiim,
69077-000, Manaus-AM, Brazil}
\affiliation{National Institute of Science and Technology for Complex Systems, 3000, Japiim,
69077-000, Manaus-AM, Brazil}

\date{\today}

\begin{abstract}
In this paper, we report mean-field and effective-field renormalization group calculations on the isotropic Heisenberg 
antiferromagnetic model under a longitudinal magnetic field. As is already known, these methods, denoted by MFRG and EFRG, 
are based on the comparison of two clusters of different sizes, each of them trying to mimic certain Bravais lattice. Our 
attention has been on the obtention of the critical frontier in the plane of temperature versus magnetic field, for the 
simple cubic and the body-centered cubic  lattices.  We used clusters with $N=1,2,4$ spins so as to implement MFRG-12, 
EFRG-12 and EFRG-24 numerical equations. Consequently, the resulting frontier lines show that EFRG approach overcomes 
the MFRG problems when clusters of larger sizes are considered.

\end{abstract}

\maketitle

\section{Introduction\protect\nolinebreak}%

\begin{figure}[htbp]
\centering
\includegraphics[width= 7.0 cm,height= 7.5 cm]{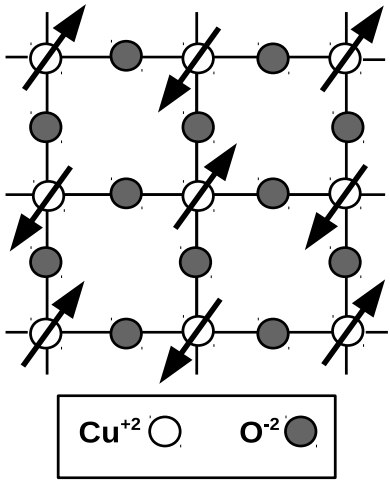}
\caption{ $\bf{CuO_{2}}$ sheet of the undoped material $\bf{La_{2}CuO_{4}}$,  showing the antiferromagnetic alignment of  $\bf{Cu^{+2}}$ spins.}
\label{figure1}
\end{figure}

The Heisenberg model has its name due an early article on the theory of ferromagnetism written by W. Heisenberg \cite{heisenberg1928}. 
Its Hamiltonian  can arise as an exchange interaction between electrons on different sites or atoms. This  was early explained by Van Vleck \cite{vanvleck}, based on Dirac's arguments \cite{dirac1926,dirac1929}. Accordingly, it can be deduced by considering a lattice of dynamical 
electrons, one per site, being $t$ the hopping amplitude between two neighboring electrons, having a strong repulsion $U$ when a site is 
doubly occupied. So, if  $U/t$ is large, the electrons prefer to occupy different sites, and transitions in which an electron on a site $i$ 
hops to a nearest occupied site $j$, are allowed only when both have  anti-parallel spins. The isotropic Hubbard Model 
\cite{hubbard1964,barnes1991} can describe this situation, and by a first-order perturbation it leads to an effective Hamiltonian given by:

\begin{equation}
\displaystyle H_{eff} = J \sum_{(ij)} \vec{S}_{i}.\vec{S}_{j} + C,
\end{equation}%
where $J=4t^{2}/U$ is positive, $\vec{S}_{i}$ is the spin operator of the electron on the site $i$, and $C$ is an additive constant. This 
is the quantum  antiferromagnetic Heisenberg model, which  has attracted much interest on account of its relation to high-temperature 
superconductivity \cite{manousakis}. Anderson suggested that two dimensional spin-$1/2$ Heisenberg antiferromagnets can be used to model 
"precursor insulators" of the high temperature superconductors \cite{Anderson1987}. For instance, the undoped material $\bf{La_{2}CuO_{4}}$ 
is an antiferromagnetic insulator \cite{bir90}, consisting of sheets of $\bf{CuO_{2}}$ separated from each other by intermediate nonmagnetic 
layers \cite{Shirane1987,Endoh1988,Aeppli1989}. The antiferromagnetic alignment of the $\bf{Cu^{+2}}$ spins can be observed in 
Fig.\ref{figure1}. In order to transform this insulator to a superconducting metal, some of the $\bf{La^{+3}}$ ions are replaced by 
dopants that prefer a $+2$ charge state \cite{bednorz1986}.

From the experimental point of view, there is also an interest in studying Heisenberg antiferromagnets under an external magnetic field. For 
instance, Neutron diffraction studies for the  spin-$1/2$ case, have obtained the saturation field $H_{c}(T)$, at which the long-range magnetic 
order is destroyed, for different spin-$1/2$ families on the square lattice \cite{coomer}. On the other hand, in the 3D case, the whole phase 
diagram in the $H-T$ plane, has not been discussed in the literature, as far as we know. For lower temperatures, a variational treatment 
showed a critical behavior of the form $H/{zJ} = 1-A(T/J)^{3/2}$, on the cubic lattice ($z=6$) \cite{falk}. Furthermore, for the bcc lattice, 
high-temperature series expansions, at low magnetic fields, have analyzed the critical behavior of the quantum metamagnetic spin-$1/2$ 
Heisenberg model, where the Neel temperature was found to depend on the magnetic field as $T_{N}(H) = T_{N}(0) [1-B(H/{zJ})^{2.08}]$ 
\cite{pan}.

In this model synchrotron radiation measurements to observed the Dzyaloshinskii-Moriya interaction in a weak ferromagnetic \cite{dmitrienko2014} 
critical and reentrant behavior \cite{minos2014} and anomaly at low temperature \cite{minos2015} in the phase diagram it has been studied. 
Recently, magnets with strong spin-orbit coupling are currently in the focus of intense research, a primary motivation being the search for 
novel phases beyond the territory of the spin-isotropic Heisenberg model \cite{jackeli2009,chaloupka2010,nussinov2015}. The paradigmatic 
Heisenberg-Kitaev model has been studied to understand the strong spin-orbit coupling in magnets in an attempt to search for new quantum 
phases beyond the territory traditionally described by the Heisenberg spin-isotropic model \cite{lukas2016}.

\vspace{0.5cm}

The study of three-dimensional spin-$1/2$ antiferromagnetic Heisenberg model under a magnetic field have been studied throughout the 
literature. One of the first works of this kind is that of Bublitz and de Sousa \cite{bublitz}, in which the criticality of the quantum 
Heisenberg antiferromagnet under a magnetic field was described for the simple cubic lattice (sc). They used a cluster with two spins with 
nearest-neighbor interactions (EFT-2) so as to obtain the phase diagram in the plane of field versus temperature. For the body-centered 
cubic lattice (bcc) these authors \cite{bublitzbcc} also obtained this type of phase diagram within mean-field approach (with 
clusters of two spins) and compared the critical frontier with points obtained from high-temperature series expansions calculated for 
lower fields \cite{pan}. More recently, this phase diagram was obtained numerically for clusters with four spins (EFT-4) 
\cite{minos2012} and compared with experimental results \cite{genoud1995} (compound $BaCuO_{2}$) in the low-field regime with a critical 
behavior of the type $T(H)/T(0)\simeq1+aH^{2}+bH^{3}$.

Accordingly,  our aim  is to improve the knowledge of the phase diagram of the antiferromagnetic Heisenberg model by combining 
effective-field (EFT) approach with the renormalization group (RG) method (the latter mixing technique is denoted as EFRG) 
\cite{plasfig}. We believe that this should better the accuracy of the calculations. In fact, these ideas come from Indekeu's paper 
\cite{indekeu}. Something similar was applied by Slotte in his doctoral thesis for the Ising model \cite{slotte} by mixing MF 
and RG, which is the Mean-Field Renormalization Group method (MFRG) (see also Reference \cite{plasfig}). Then, Li and Yang 
implemented for the first time the EFRG for the diluted Ising model, to the best of our knowledge \cite{liyang}.

These two methods have been applied to a variety of magnetic spin models in obtaining of the critical properties of classical 
XY and Heisenberg models \cite{sadeghi2007}, critical behavior of two- and three-dimensional ferromagnetic and antiferromagnetic 
spin-ice systems \cite{angel2001}, quenched site-dilute ising models \cite{bobak1994} and to obtain the reduced critical 
temperature and exponents $\nu$ for bi- and threedimensional lattices \cite{albuquerque2009}.

In this work our interest is in improving the study of three-dimensional spin-$1/2$ antiferromagnetic Heisenberg model 
under a magnetic field by using EFRG and mean-field renormalization group MFRG. In the following section we will explain the 
formalism for treating the present model in the simple cubic (sc) and the body-centered cubic (bcc) lattices. In section 
III we present the results and in the IV section the conclusions. 

\section{Model and Formalism}

In this work we treat  the isotropic antiferromagnetic Heisenberg model with  nearest-neighbors ($nn$) in a longitudinal magnetic 
field divided into two equivalent interpenetrating sub-lattices $A$ e $B$, that is represented by the following Hamiltonian

\begin{equation}
H=J\sum\limits_{<i,j>}\left(\sigma _{i}^{x}\sigma_{j}^{x}+\sigma _{i}^{y}\sigma_{j}^{y}+\sigma _{i}^{z}\sigma_{j}^{z}\right)-H\sum\limits_{i}\sigma _{i}^{z},
\label{1}
\end{equation}%
where $J$ stands for the antiferromagnetic exchange interaction, $\langle i,j\rangle$ denotes the sum over all pairs of nearest-neighbor 
spins ($z$) on a  lattice (here we work in three-dimensional lattices, $z=6$ and $z=8$) and $\sigma _{i}^{\nu}$ is the $\nu (=x,y,z)$ 
component of the spin-$1/2$ Pauli operator at site $i$.

The competition between the antiferromagnetic exchange and the magnetic field present interesting properties in the corresponding 
phase diagram. Particularly, the model given in Eq.(\ref{1}) maintains an antiferromagnetic  phase (AF) even in the presence of a 
field (up to some critical value).

\subsection{Clusters in effective-field theory}

Before we justify the hypothesis of homogeneity in order to treat the model (\ref{1}) in three-dimensional lattices, we consider 
a simple example of one-spin $N=1$, two-spins $N^{\prime}=2$ and four-spins $N^{\prime \prime}=4$ clusters in the effective-field 
theory (which we shall call here by acronyms EFT-1, EFT-2 and EFT-4), so the respective Hamiltonians are given by

\begin{equation}
\mathcal{H}^{(1)}=\left(J\sum\limits_{\delta_{1}}\sigma_{(1+\delta_{1})}^{z}-H\right)\sigma_{1}^{z},
\label{1a}
\end{equation}

\begin{eqnarray}
\mathcal{H}^{(2)}=J^{\prime}\overrightarrow{\sigma}_{1}\cdot\overrightarrow{\sigma}_{2}
+\sum_{i=1}^{2}\sigma_{i}^{z}\left(J^{\prime}\sum\limits_{\delta_{2}}\sigma_{(1+\delta_{2})}^{z}-H^{\prime}\right)
\label{2}
\end{eqnarray}%
and 
\begin{eqnarray}
\mathcal{H}^{(4)}=  && \nonumber 
\\
J^{\prime\prime}\left(\overrightarrow{\sigma}_{1}\cdot\overrightarrow{\sigma}_{2}
+\overrightarrow{\sigma}_{2}\cdot\overrightarrow{\sigma}_{3}+\overrightarrow{\sigma}_{3}\cdot\overrightarrow{\sigma}_{4}
+\overrightarrow{\sigma}_{4}\cdot\overrightarrow{\sigma}_{1}\right) \nonumber
\\ 
+\sum_{i=1}^{4}\sigma_{i}^{z}\left(J^{\prime \prime}\sum\limits_{\delta_{4}}\sigma_{(1+\delta_{4})}^{z}-H^{\prime \prime}\right)
\label{3}
\end{eqnarray}%
where $(J\sum\limits_{\delta_{1}}\sigma_{(1+\delta_{1})}^{z}-H)$, $(J^{\prime}\sum\limits_{\delta_{2}}\sigma_{(1+\delta_{2})}^{z}-H^{\prime})$ and $(J^{\prime\prime}\sum\limits_{\delta_{4}}\sigma_{(1+\delta_{4})}^{z}-H^{\prime\prime})$ 
denoting an elementary vector of the lattice one-spin $N=1$, two-spin $N^{\prime}=2$ and four-spin $N^{\prime \prime}=4$ 
cluster respectively.

The average magnetization $m_{1A}=\langle \textbf{Tr}\sigma_{1A}e^{-\beta\mathcal{H}^{(1)}} / \textbf{Tr} e^{-\beta\mathcal{H}^{(1)}}\rangle$, $m^{\prime}_{2A}=\langle \textbf{Tr}\sigma_{2A}e^{-\beta\mathcal{H}^{(2)}} / \textbf{Tr} e^{-\beta\mathcal{H}^{(2)}}\rangle$ 
and $m^{\prime\prime}_{4A}=\langle \textbf{Tr}\sigma_{4A}e^{-\beta\mathcal{H}^{(4)}} / \textbf{Tr} e^{-\beta\mathcal{H}^{(4)}}\rangle$, 
corresponding to the Hamiltonians (\ref{1a}), (\ref{2}) and (\ref{3}), are obtained through the Callen-Suzuki relation derived in Ref. 
\cite{Sa1985}, so we have:

\begin{eqnarray}
m_{1A}(K,L)=\left\langle\prod_{\delta}^{z}(\alpha_{x}+\sigma _{(1+\delta)}\beta _{x})\right\rangle && \nonumber
\\
\left.\tanh\left(L-x\right)\right\vert _{x=0},
\label{3a}
\end{eqnarray}

\begin{eqnarray}
m_{2A}(K^{\prime},L^{\prime}) = && \nonumber 
\\ \left\langle\prod_{\delta_{1}}^{z}(\alpha^{\prime}_{x}+\sigma _{(1+\delta_{1})}^{z}\beta^{\prime}_{x}) \prod_{\delta_{2}}^{z-1}(\alpha^{\prime}_{y}+\sigma_{(2+\delta_{2})}^{z}\beta^{\prime}_{y})\right\rangle \cdot \nonumber 
\\
\left.G(x,y)\right\vert_{x=y=0}
\label{4}
\end{eqnarray}%
and 
\begin{eqnarray}
m_{4A}(K^{\prime \prime},L^{\prime \prime}) = 
\left\langle\prod_{\delta_{1}}^{z} \Phi^{\prime \prime}_{1} \prod_{\delta_{2}}^{z-1} \Phi^{\prime \prime}_{2} \prod_{\delta_{3}}^{z-2}\Phi^{\prime \prime}_{3} \prod_{\delta_{4}}^{z-3}\Phi^{\prime \prime}_{4}\right\rangle \cdot  && \nonumber
\\ 
\left. M(x,y,w,z)\right\vert_{x=y=w=z=0},
\label{5}
\end{eqnarray}
with%
\begin{equation}
G(x,y)=\frac{\sinh(A)+\frac{x-y}{W}\exp(2K)\sinh(W)}{\cosh(A)+\exp(2K)\cosh(W)}
\label{6}
\end{equation}%
where $K=\beta J$, $L=\beta H$, $A=2KL-x-y$, $W=\sqrt{(x-y)^{2}+4K^{2}}$, $\alpha_{x}=\cosh (KD_{x})$, $\beta _{x}=\sinh (KD_{x})$, $\alpha^{\prime}_{\nu}=\cosh (K^{\prime}D_{\nu})$, $\beta^{\prime}_{\nu}=\sinh(K^{\prime}D_{\nu})$ with $\nu=x,y$ and 
$D_{\nu}=\partial/\partial\nu$, and $\Phi^{\prime\prime}_{1}=(\alpha^{\prime \prime}_{x}+\sigma_{(1+\delta_{1})}^{z}\beta^{\prime \prime}_{x})$, $\Phi^{\prime\prime}_{2}=(\alpha^{\prime\prime}_{y}+\sigma_{(2+\delta_{2})}^{z}\beta^{\prime \prime}_{y})$, $\Phi^{\prime\prime}_{3}=(\alpha^{\prime \prime}_{\lambda}+\sigma_{(3+\delta_{3})}^{z}\beta^{\prime \prime}_{\lambda})$, $\Phi^{\prime\prime}_{4}=(\alpha^{\prime \prime}_{\lambda}+\sigma_{(4+\delta_{4})}^{z}\beta^{\prime \prime}_{\lambda})$, $\alpha^{\prime \prime}_{\lambda}=\cosh K^{\prime \prime}(D_{x}+D_{y}+D_{w}+D_{z})$ e $\beta^{\prime \prime}_{\lambda}=\sinh K^{\prime \prime}(D_{x}+D_{y}+D_{w}+D_{z})$, with $\lambda=x,y,w,z$ and $D_{\lambda}=\partial/\partial\lambda$. $M(x,y,w,z)$ is obtained  by numerical 
diagonalization.

We recall that Eqs. (\ref{3a}), (\ref{4}) and (\ref{5}) are approximate and will be used here as the basis of the present formalism. In 
the Ising case, where spin operators have only the component $z$, both equations are exact, and have been studied in Ref. \cite{Kincaid1975,Wentworth1993,Moreira2002,Neto2004}. Here, we use a decoupling procedure that ignores all high-order spin correlations 
on both right-hand sides of Eqs. (\ref{3a}), (\ref{4}) and (\ref{5}), i.e., 

\begin{equation} 
\langle \sigma_{i}^{z}\sigma_{j}^{z}\cdots \sigma_{n}^{z}\rangle=\langle \sigma_{i}^{z}\rangle \langle\sigma_{j}^{z}\rangle\cdots\langle\sigma_{n}^{z}\rangle,
\label{7}
\end{equation}%
with $i\neq j\neq \cdots n$, and we adopt $b_{\eta}=\langle\sigma_{i\eta}\rangle$, $b^{\prime}_{\eta}=\langle\sigma^{\prime}_{i\eta}\rangle$ 
and $b^{\prime \prime}_{\eta}=\langle\sigma^{\prime \prime}_{i\eta}\rangle$ for the clusters with $N=1$, $N^{\prime}=2$ and 
$N^{\prime\prime}=4$ spins, respectively (where $b$, $b^{\prime}$ and $b^{\prime \prime}$ are the symmetric breaking effective fields) with 
($\eta=A$ and $B$) denotes the corresponding sub-lattice. 

\begin{figure}[htbp]
\centering
\includegraphics[width= 7.0 cm,height= 5.5 cm]{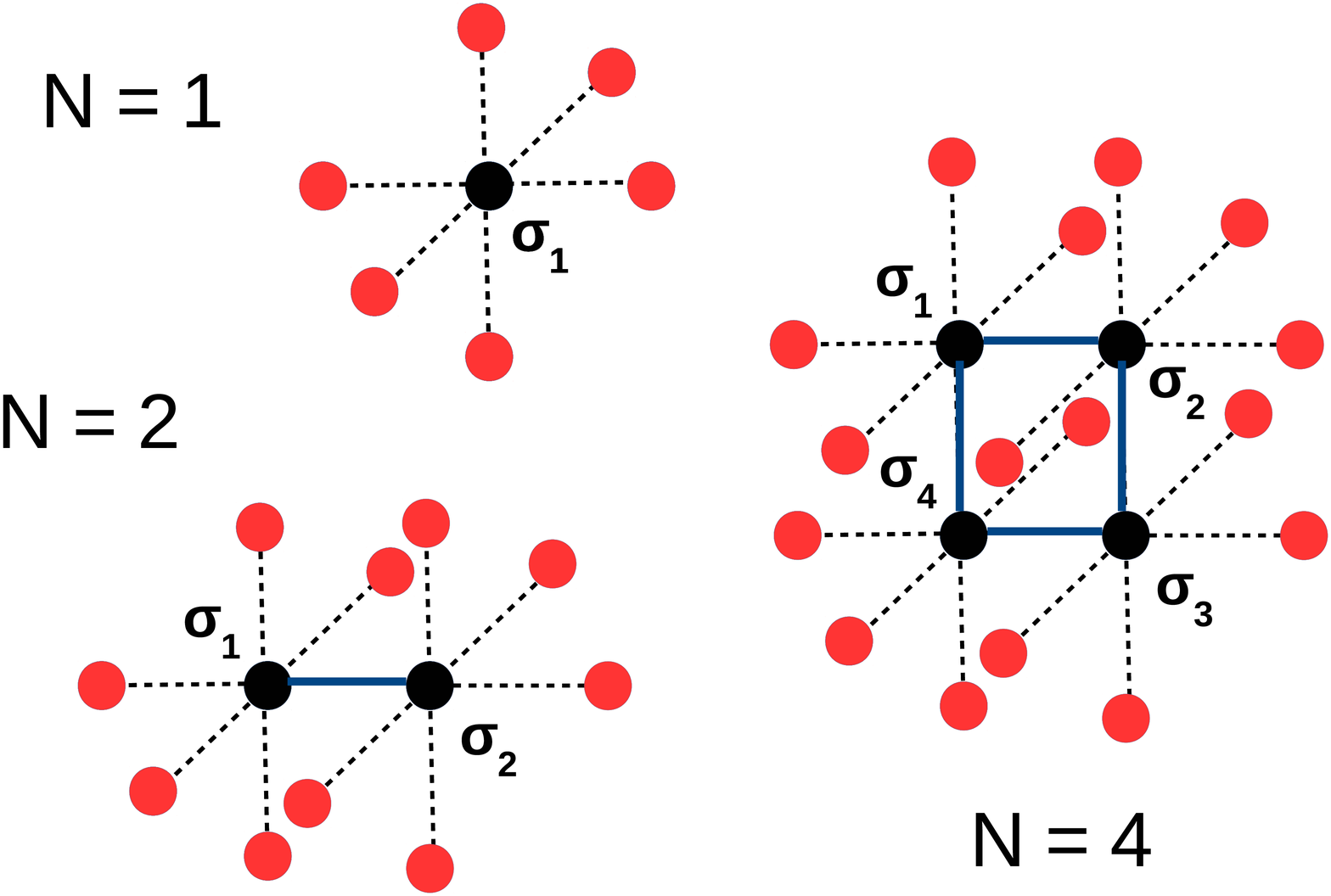}
\caption{Clusters with $N=1,2,4$ spins used in the effective-field technique for the cubic lattice considering nearest-neighbor interactions.}
\label{figure2}
\end{figure}

\begin{figure}[htbp]
\centering
\includegraphics[width= 7.0cm,height= 5.5 cm]{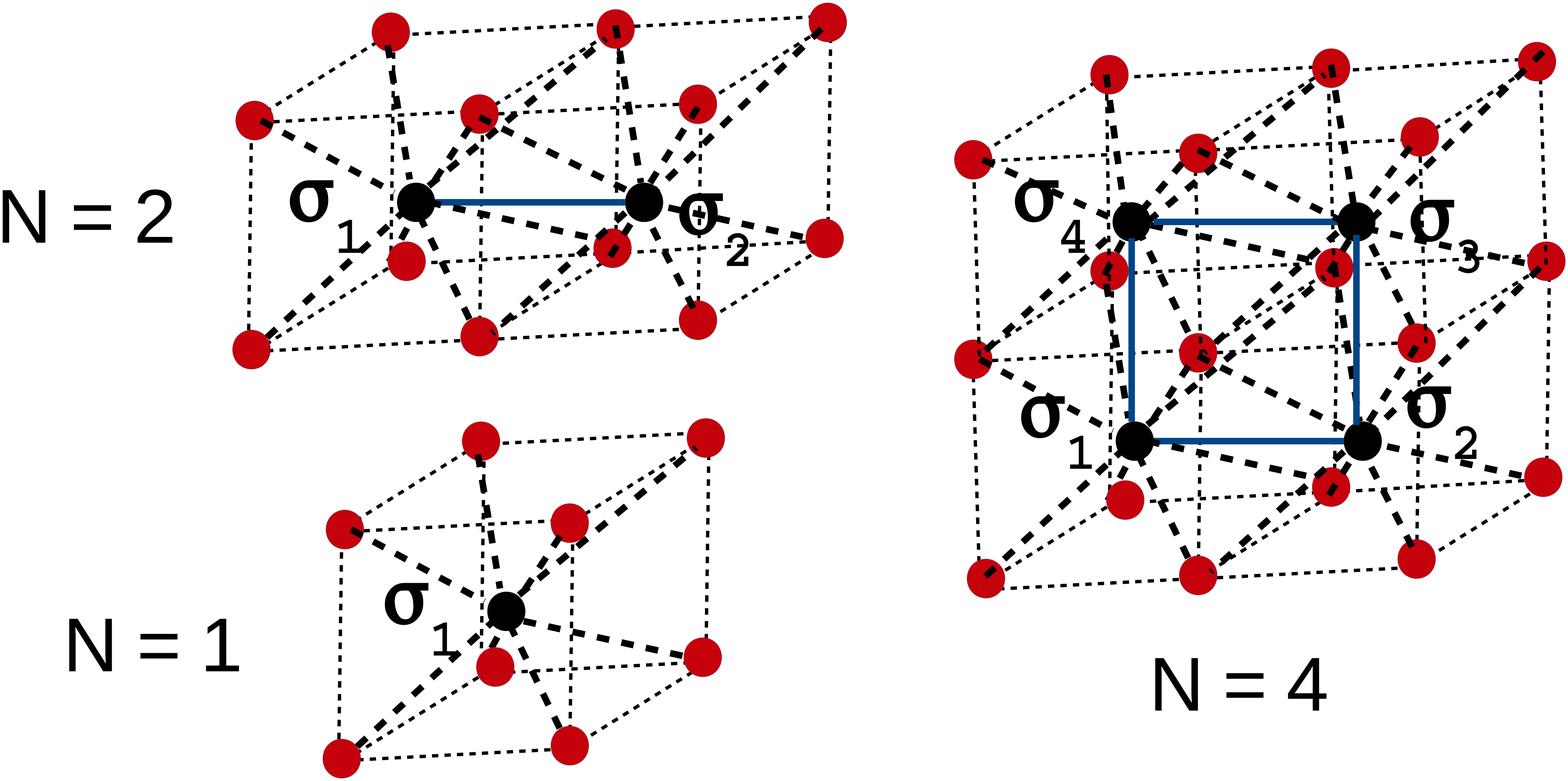}
\caption{ Clusters with $N=1,2,4$ spins for the bcc lattice with nearest-neighbor interactions.}
\label{figure3}
\end{figure}

Before applying the decoupling, we must observe that in Eqs. (\ref{3a}), (\ref{4}) and (\ref{5}) sites $1$ $2$ and $4$ of the one-spin, 
two-spins and four-spins clusters may exhibit a set of common nearest-neighbors sites (see Figs. \ref{figure2} and \ref{figure3}). Thus, 
Eqs. (\ref{3a}), (\ref{4}) and (\ref{5}) can be re-casted in the following form:

\begin{equation}
m_{1A}(K,L)=\left(\alpha_{x}+b_{B}\beta_{x}\right)^{z}\left.\tanh\left(L-x\right)\right\vert_{x=0};
\label{7a}
\end{equation}

\begin{eqnarray}
m_{2A}(K^{\prime},L^{\prime})= && \nonumber
\\
\left(\alpha_{x}^{\prime}+b_{B}^{\prime}\beta_{x}^{\prime}\right)^{z-1}\left(\alpha_{y}^{\prime}+b_{A}^{\prime}\beta_{y}^{\prime}\right)^{z-1} \left.G(x,y)\right\vert_{x=y=0},
\label{8}
\end{eqnarray}%
and 
\begin{eqnarray}
m_{4A}(K^{\prime \prime},L^{\prime \prime})=\left(\alpha_{x}^{\prime \prime}+b_{B}^{\prime \prime}\beta_{x}^{\prime \prime}\right)^{z-2}\left(\alpha_{y}^{\prime\prime}+b_{A}^{\prime\prime}\beta_{y}^{\prime\prime}\right)^{z-2} && \nonumber
\\
\left(\alpha_{w}^{\prime\prime}+b_{B}^{\prime\prime}\beta_{w}^{\prime\prime}\right)^{z-2}
\left(\alpha_{z}^{\prime\prime}+b_{A}^{\prime\prime}\beta_{z}^{\prime\prime}\right)^{z-2} \cdot && \nonumber
\\
\left.L(x,y,w,z)\right\vert _{x=y=w=z=0}.
\label{9}
\end{eqnarray}%

This system has two distinct sub-lattice, which in the ordered phase (AF) have different magnetizations (and symmetry-breaking fields). 
The order parameter of the two sub-lattice are $m_{s}=(m_{A}-m_{B})/2$ and $m=(m_{A}+m_{B})/2$, as the staggered and the uniform 
magnetizatons, respectively. The expansion of the right-hand side of (\ref{7a}), (\ref{8}) and (\ref{9}) in powers of the parameters 
$b=(b_{A}+b_{B})/2$, $b_{s}=(b_{A}-b_{B})/2$, $b^{\prime}=(b^{\prime}_{A}+b^{\prime}_{B})/2$, 
$b^{\prime}_{s}=(b^{\prime}_{A}-b^{\prime}_{B})/2$, $b^{\prime \prime}=(b^{\prime \prime}_{A}+b^{\prime \prime}_{B})/2$, $b^{\prime \prime}_{s}=(b^{\prime \prime}_{A}-b^{\prime \prime}_{B})/2$, in first order in $b_{s}$, $b^{\prime}_{s}$ and $b^{\prime \prime}_{s}$ 
are given by

\begin{equation}
m_{1s}(K,L)\simeq A_{1s}(K,H,b)b_{s},
\label{10}
\end{equation}

\begin{equation}
m_{1}(K,L)\simeq A_{1}(K,H,b),
\label{11}
\end{equation}

\begin{equation}
m_{2s}(K^{\prime},L^{\prime})\simeq A_{2s}(K^{\prime},L^{\prime},b^{\prime})b^{\prime}_{s},
\label{11a}
\end{equation}

\begin{equation}
m_{2}(K^{\prime},L^{\prime})\simeq A_{2}(K^{\prime},L^{\prime},b^{\prime}),
\label{11b}
\end{equation}

\begin{equation}
m_{4s}(K^{\prime \prime},L^{\prime \prime})\simeq A_{4s}(K^{\prime \prime},L^{\prime \prime},b^{\prime \prime})b^{\prime \prime}_{s},
\label{11c}
\end{equation}%
and 
\begin{equation}
m_{4}(K^{\prime \prime},L^{\prime \prime})\simeq A_{2}(K^{\prime \prime},L^{\prime \prime},b^{\prime \prime}),
\label{11d}
\end{equation}

\vspace{0.1cm}

\vspace{0.1cm}

The critical frontier line which separates the antiferromagnetic and the disordered phase is obtained by solving numerically 
with cluster one-spin, that constitutes the EFT-1, solving Eqs. (\ref{10}) and (\ref{11}). Analogously to clusters two- (the 
Eqs. (\ref{11a}) and (\ref{11b}), EFT-2) and four-spins (\ref{11c}) and (\ref{11d}) the EFT-4, respectively. These last two cases have 
already been analyzed by Neto \textit{et al.} \cite{minos2012}. In other words, by these equations we can determine numerically the 
dependence of the critical field on the temperature. The effective-field theory in larger clusters has been studied in Ising model 
\cite{akinci2015}.

\subsection{Clusters in mean-field approximation}

We will use the same idea of the calculations in EFT-1 and EFT-2, only now for the approximations with one- and two-spins by the 
mean-field approximation (MFA) which we will call here of MFA-1 and MFA-2, respectively. Using a variational method based on Peierls-Bogoliubov inequality (which is based on arguments of convexity, see \cite{falk1970}), which can be formally written for any classical 
and quantum system as

\begin{equation}
F\left(\mathcal{H}\right)\leq F_{0}\left(\mathcal{H}_{0}\right)+\left\langle
\mathcal{H}-\mathcal{H}_{0}\right\rangle_0\equiv\Phi(\eta), 
\label{12}
\end{equation}%
where $F$ and $F_{0}$ are free energies associated with two systems defined by the Hamiltonians $\mathcal{H}$ and $\mathcal{H}_{0}(\eta)$, 
respectively, the thermal average $\left\langle...\right\rangle_0$ should be taken in relation to the canonical distribution associated 
with the trial Hamiltonian $\mathcal{H}_{0}(\eta)$, with $\eta$ standing for the variational parameters. The approximated free energy 
$F$ is given by the minimum of $\Phi(\eta)$ with respect to $\eta$, i.e. $F\equiv\Phi_{min}(\eta)$.

The trial Hamiltonian $\mathcal{H}^{1}_{0}$ is chosen as free spins, distributed in two different sub-lattices $A$ and $B$. We then have for 
$N=1$

\begin{equation}
\mathcal{H}^{1}_{0}=-\eta_A\sum_{i\subset A}\sigma _{i}^{z}-\eta_B\sum_{i\subset B}\sigma _{i}^{z}-H\sum_{i\subset A,B}\sigma_{i}^{z},
\label{13}
\end{equation}%
where $\eta_A$ and $\eta_B$ are two variational parameters.

To compute the right hand side of Eq. (\ref{12}) and, after minimizing $\Phi(\gamma)$, the variational parameters $\eta_A$ and
$\eta_B$ can be written as a function of the sub-lattice magnetizations $m_{A}$ and $m_{B}$. The approximated mean-field Helmholtz 
free energy per spin, $f=\Phi/N$, can thus be written as

\begin{eqnarray}
f=-\frac{t}{2}\ln\left\lbrace4\cosh(L-zKm_{A})\cosh(L-zKm_{B})\right\rbrace && \nonumber 
\\
+\frac{z}{2}m_{A}m_{B},
\label{14}
\end{eqnarray}%
and the corresponding sub-lattice magnetizations $m_{A}$ and $m_{B}$ given by

\begin{equation}
m_{A}=\tanh(L-zKm_{B})
\label{15}
\end{equation}%
and
\begin{equation}
m_{B}=\tanh(L-zKm_{A}).
\label{16}
\end{equation}

For two-spins agglomerated we have written the trial Hamiltonian $\mathcal{H}^{2}_{0}$ is selected to be of similar form given by

\begin{eqnarray}
\mathcal{H}^{2}_{0}=J\sum_{s}\left(\sigma^{x}_{1s}\sigma^{x}_{2s}+\sigma^{y}_{1s}\sigma^{y}_{2s}
+\sigma^{z}_{1s}\sigma^{z}_{2s}\right) && \nonumber
\\
-\sum_{s}\eta_A\sigma_{As}-\eta_B\sigma_{Bs}, 
\label{17}
\end{eqnarray}%
where the sum extends over all pairs of spins $n_{s}=N^{\prime}/2$. Substituting Eq. (\ref{17}) in Eq. (\ref{12}), we obtain the variational 
free energy function per spins that is

\begin{eqnarray}
f =-\frac{t}{2}\ln\left\lbrace 2e^{K^{\prime}}\cosh\Delta_{1}+2e^{-K^{\prime}}\cosh\Delta_{2}\right\rbrace && \nonumber
\\
-\frac{(z-1)}{2}m_{A}m_{B},
\label{18}
\end{eqnarray}%
where $\Delta_{1}=2L^{\prime}-(z-1)K^{\prime}(m_{A}+m_{B})$ and $\Delta_{2}=\sqrt{(z-1)^{2}(m_{A}-m_{B})^{2}+4K^{\prime 2}}$ and the corresponding sub-lattice magnetizations $m_{A}$ and $m_{B}$ given by

\begin{equation}
m_{A}=\frac{e^{-K^{\prime}}\sinh\Delta_{1}+e^{K^{\prime}}\Delta_{3}\sinh\Delta_{2}}{e^{-K^{\prime}}\cosh\Delta_{1}+e^{K}\cosh\Delta_{2}}
\label{19}
\end{equation}%
and
\begin{equation}
m_{B}=\frac{e^{-K^{\prime}}\sinh\Delta_{1}-e^{K^{\prime}}\Delta_{3}\sinh\Delta_{2}}{e^{-K^{\prime}}\cosh\Delta_{1}+e^{K}\cosh\Delta_{2}},
\label{20}
\end{equation}%
with $\Delta_{3}=(z-1)(m_{A}-m_{B})/\Delta_{2}$.

Again, we can define the symmetry breaking fields and the expand of the right-hand side of (\ref{17}) and (\ref{19}) (or (\ref{18}) and (\ref{20})) we have

\begin{equation}
m_{1s}(K,L)\simeq \left[1-\tanh^{2}(L-zKb_{s})\right]zKb_{s},
\label{21}
\end{equation}

\begin{equation}
m_{1}(K,L)\simeq \tanh(L-zKb),
\label{22}
\end{equation}

\begin{eqnarray}
m_{2s}(K^{\prime},L^{\prime}) \simeq && \nonumber
\\
\frac{(z-1)b^{\prime}_{s}\sinh(2K^{\prime})}{e^{-2K^{\prime}}\cosh\left[2L^{\prime}-(z-1)K^{\prime}b^{\prime}_{s}\right]+\cosh(2K^{\prime})},
\label{23}
\end{eqnarray}%
and

\begin{eqnarray}
m_{2}(K^{\prime},L^{\prime}) \simeq && \nonumber
\\
\frac{\sinh\left[2L^{\prime}-(z-1)K^{\prime}b^{\prime}_{s}\right]}{\cosh\left[2L^{\prime}-(z-1)K^{\prime}b^{\prime}_{s}\right]+e^{2K^{\prime}}\cosh(2K^{\prime})}.
\label{24}
\end{eqnarray}

The critical frontier line which separates the antiferromagnetic and the disordered phase is obtained by solving numerically 
cluster with one-spin, that constitutes the MFA-1, solving Eqs. (\ref{21}) and (\ref{22}). Analogously to clusters two-spins the 
Eqs. (\ref{23}) and (\ref{24}) that constitutes the MFA-2, where the field-induced phase transition in the quantum Heisenberg 
antiferromagnet was obtained by Bublitz and de Sousa \cite{bublitzbcc}.

\subsection{Renormalization group}

A hypothesis in the critical region is the expansion of free energy into a power series. In some models it is not possible to 
write an expansion where the coefficients are functions of the temperature, as for example, the two-dimensional Ising model. Then, the 
hypothesis of scale is formulated in an attempt to justify the thermodynamic behavior in the critical region. This basis and the 
possibility of calculating the critical exponents were possible by the renormalization group \cite{salinas}.

Near the critical point the correlation length is much larger than the distance between the network sites. Kadanoff presented in the 
1970s \cite{kadanoff1977} that it is possible to decrease the number of degrees of freedom of the system. The general scheme of the 
renormalization group

In order to apply the renormalization group theory so as to improve the mean- and effective-field approaches in the present model, we 
use the proposal given in the paper of Indekeu, Maritan and Stella \cite{indekeu}. This consists of relating the magnetization of a 
finite cluster of $N$ spins with that of $N' (N > N')$ given by:

\begin{equation}
m_{N^{\prime}} ({K^{\prime}_{i}}, b^{\prime}, L^{\prime}) = l^{d-y_{h}} m_{N} ({K_{i}}, b, L),
\label{similaridade}
\end{equation}%
where $b (b^{\prime})$ is the field of symmetry breaking associated to the cluster of $N(N^{\prime})$ spins, $d$ the lattice dimension, $K = J/{k_{B}T}$, $L=H/{k_{B}T}$, and $y_{L}$ is magnetic critical exponent of the scaling relation $L^{\prime}= l^{y_{L}}h$, where $l=(N/N^{\prime})^{1/d}$. Then, a first-order expansion in $b(b^{\prime})$ is done for the order parameter $m_{N}(m_{N^{\prime}})$ so as to equate the first-order terms:

\begin{equation} 
A_{N^{\prime}}({K^{\prime}_{i}}, L^{\prime}) b^{\prime} = l^{d-y_{L}} A_{N}({K_{i}}, L) b.
\end{equation}%
The difficulty of solving the above equation rests on the unkowledge of the critical exponent $y_{L}$. This is where Indekeu \textit{et al.} 
proposed that the fields $b^{\prime}$ and $b$ have the same hypothesis of similarity  as Eq. (\ref{similaridade}), so $b^{\prime}= l^{d-y_{L}}$, which leads to 

\begin{equation}
A_{N^{\prime}}({K^{\prime}_{i}}, h^{\prime})= A_{N}({K_{i}}, L). 
\end{equation}
This is the basic idea of the MFRG and EFRG that considers that in the criticality the fixed point is reached when $K^{\prime}_{i} = K_{i} = K^{*}_{i}$ and $L^{\prime}_{i} = L_{i} = L^{*}_{i}$. The effect of the critical behavior on the surface will not be considered \cite{indekeu1987}.

\subsection{EFRG-12 approach}

Accordingly, we may apply this concept to the present model so as to obtain the critical frontier in the $H-T$ plane for EFRG schemes for clusters with $N=1$ and $N^{\prime}=2$, and for clusters with $N^{\prime}= 2$ and $N^{\prime\prime}=4$, denoted by EFRG-12 and EFRG-24, respectively. Thus, 
we have for the EFRG-12 approach for clusters of one and two spins, the fixed critical point 
$K^{\prime}=K=K^{*}=1/T_{N}(H)$ is obtained from Eqs. (\ref{10}) and (\ref{11a}):

\begin{equation}
A_{1s}(K^{*}, L, b) = A_{2s}(K^{*}, L^{\prime}, b^{\prime}),
\label{eqs12}
\end{equation}%
where this critical condition now depends on the non-critical variables $b$ and $b^{\prime}$, and a reasonable choice for the size 
dependence of these must be given in order to proceed. This choice was proposed by Plascak and Sá Barreto \cite{plascak1986}, that 
postulate an identity between the symmetry breaking field and uniform magnetization of each cluster, i.e., 

\begin{equation}
m(T)=b=b^{\prime}
\label{eqs12a}
\end{equation}%
and
\begin{equation}
A_{1}(K, L, b) = A_{2}(K^{\prime}, L^{\prime}, b^{\prime}).
\label{eqs12b}
\end{equation}

We can simultaneously solve the set of three equations (\ref{eqs12}-\ref{eqs12b}) with $L^{\prime}=L\equiv h/T_{N}$ $(h=H/J)$, and obtain 
the values of $T_{N}$ and $b=b^{\prime}$ for each value of the intensity of the external field $H$ on a sc and bcc lattices.

\subsection{EFRG-24 approach}
In a very similar way, we can do for the EFRG-24 approach for clusters of two and four spins is obtained from Eqs. 
(\ref{11a}) and (\ref{11c}):

\begin{equation}
A_{2s}(K^{*}, L^{\prime}, b^{\prime}) = A_{4s}(K^{*}, L^{\prime \prime}, b^{\prime \prime}),
\label{eqs12c}
\end{equation}%
where this critical condition now depends on the non-critical variables $b^{\prime}$ and $b^{\prime \prime}$,

\begin{equation}
m(T)=b^{\prime}=b^{\prime \prime}
\label{eqs12d}
\end{equation}%
and
\begin{equation}
A_{2}(K^{\prime}, L^{\prime}, b^{\prime}) = A_{4}(K^{\prime \prime}, L^{\prime \prime}, b^{\prime \prime}).
\label{eqs12e}
\end{equation}

In the next section we will apply simultaneously the set of three equations (\ref{eqs12c}-\ref{eqs12e}) 
with $L^{\prime}=L\equiv h/T_{N}$ $(h=H/J)$. In Figs. (\ref{figure2}) and (\ref{figure3}) can be observed the respective cluster 
sizes that will be used for each of them.

\subsection{MFRG-12 approach}

Furthermore, for the sake of completeness, we also did calculations within the MFRG-12. Assuming, once again, that the fields 
of symmetry breaking associated with their respective magnetizations are scaled in the same way, we can obtain from Eqs. (\ref{21}) and 
(\ref{23}) the relation between $K$ and $K^{\prime}$ is given by

\begin{eqnarray}
\left[1-\tanh^{2}(L-zKb_{s})\right]zKb_{s}= && \nonumber 
\\
\frac{(z-1)b^{\prime}_{s}\sinh(2K^{\prime})}{e^{-2K^{\prime}}\cosh\left[2L^{\prime}-(z-1)K^{\prime}b^{\prime}_{s}\right]+\cosh(2K^{\prime})}.
\label{25}
\end{eqnarray}

We note that for null magnetic fields $b$ and $b^{\prime}$ are different from zero and therefore from Eq. (\ref{25}) we can not obtain the 
fixed point $K=K^{\prime}=K_{N}$ because we do not know $b$ and $b^{\prime}$, so we have three variables to be determined for only one 
equation. However, there are two distinct ways in which to work around this problem in the literature.

The first is that fields $b$ and $b^{\prime}$ are obtained via the mean-field approximation, \textit{i.e.}, $m_{1}=b$ and $m_{2}=b^{\prime}$ 
(equating the Eqs. (\ref{22}) and (\ref{24})) the $b$ and $b^{\prime}$, respectively). The other would be based on invariance of scale, where 
the Eqs. (\ref{22}) and (\ref{24}) are equivalents, \textit{i.e.}, 

\begin{eqnarray}
\tanh(L-zK_{N}b)\equiv && \nonumber
\\
\frac{\sinh\left[2L^{\prime}-(z-1)K_{N}b^{\prime}_{s}\right]}{\cosh\left[2L^{\prime}-(z-1)K_{N}b^{\prime}_{s}\right]+e^{2K_{N}}\cosh(2K_{N})},
\label{26}
\end{eqnarray}%
which we shall here denote by invariance of scale method. There is not that much difference in both methods, so we will use this latter 
method in our results.

\section{Results}

For quantum Heisenberg model and in particular its phase diagram within the EFRG framework can be deduced from the Eqs. (\ref{eqs12}-\ref{eqs12e}). 
For the Ising case some works have been published \cite{Kincaid1975,Wentworth1993,Moreira2002,Neto2004}. The phase diagram in the $H-T$ plane 
comprises a metamagnet or field-induced antiferromagnetic phase ($m_{s}\neq0$) at low fields and a paramagnetic phase ($m_{s}=0$) at higher fields. Furthermore, in the absence of next-nearest-neighbor interactions (nnn), there is only a critical line which separates the paramagnetic 
(\textbf{P}) and antiferromagnetic (\textbf{AF}) phases by a continuous phase transition (note what is the magnetic phase for $\alpha =0$ in 
Ref. \cite{oliveira}, which treats the square lattice).

\begin{figure}[htbp]
\centering
\includegraphics[width=8.0cm,height=7.0cm]{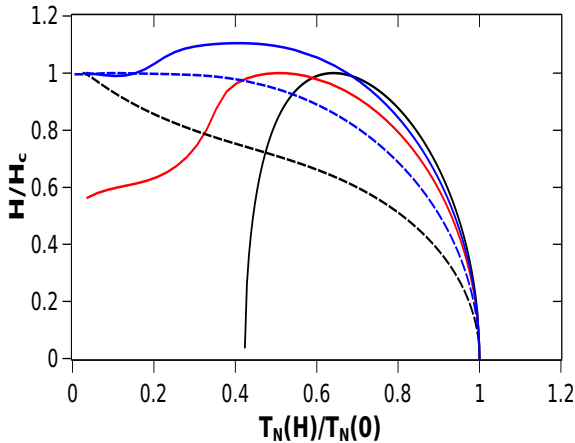}
\caption{The dependence of the reduced magnetic field $H/H_{c}$ as a function of the reduced temperature $T_{N}(H)/T_{N}(0)$ of the quantum isotropic spin-$1/2$ Heisenberg antiferromagnetic model on the sc lattice $(z=6)$. The black, red and blue continuos lines and black dashed scheme are MFRG-12, EFRG-12, EFRG-24 and MFA-2 schemes, respectively, are our contributions. The blue dashed line is EFT-4 scheme \cite{minos2012}.} 
\label{figure4}
\end{figure}

In Fig. (\ref{figure4}) we observe the frontier that divides the antiferromagnetic and the paramagnetic phase for the sc lattice, which was 
obtained numerically by three different approaching methods, namely, MFRG-12, EFRG-12 and EFRG-24. The MFRG-12 frontier line shows a sharp 
reentrant behavior that is progressively corrected by EFRG.

\vspace{0.05cm}

At high-temperature the entropy is the predominant factor, in the appearance of the reentrant phenomenon, and the system is then in the disordered P phase but with an AF bias due the applied fields \cite{minos2013}. This behavior tends to decrease as we improve the approach. This is made clear in the sc lattice, see Fig. (\ref{figure4}). In the MFRG-12 approach, this behavior extends practically to all values of the magnetic field. In this lattice there is an inflection point around $\left(T_{N}/T_{N}(0)\right)\cong 0.209$ in the EFRG-12 and EFRG-24 approaches. The maximum point of each curve occurs at values: $(2.600,3.255)$ MFRG-12, $(2.050,3.839)$ EFRG-12 and $(1.622,3.985)$ EFRG-24, with non-normalized values presented.     

This reentrant zone is for temperatures in the interval $ 0.415 < T_{N}/T_{N}(0) < 1.0$, where $T_{N}(H=0) = 4.06 J/K_{B}$. For EFRG-12 and EFRG-24, these intervals are  $ 0.035 < T_{N}/T_{N}(0) < 0.92$, for fields between $ 0.55 < H/H_{c} < 0.95$ and $ 0.035 < T_{N}/T_{N}(0) < 0.71$, for fields between $ 1.0 < H/H_{c} < 1.1$, respectively. The temperature critical to the null field for EFRG-12 is $T_{N}(H=0) = 4.09 J/K_{B}$ and $T_{N}(H=0) = 4.07 J/K_{B}$ (EFRG-24).

Furthermore, our critical temperature at zero field is in agreement with the EFRG-12 value $T_{N}(H=0) \simeq 4.09$ (in $J/k_{B}$ units) obtained by de Sousa and Araujo \cite{araujo}. This numerical results obtained is also compared with other different methods, such as spin wave theory $T_{N} \simeq 4.42$ \cite{wei1993}, Green's function method $T_{N}(H=0) \simeq 3.62$ \cite{takahashi1989}, Monte Carlo study $T_{N}(H=0) \simeq 3.54$ \cite{he1993}, high-temperature expansion $T_{N}(H=0) \simeq 3.59$ \cite{rush1967}, variational cumulat expansion $T_{N}(H=0) \simeq 4.59$ \cite{li1995}, EFT-2 $T_{N}(H=0) \simeq 4.95$ and EFT-4 $T_{N}(H=0) \simeq 4.81$ \cite{minos2012}.

\vspace{0.05cm}
\begin{figure}[htbp]
\centering
\includegraphics[width=8.0cm,height=7.0cm]{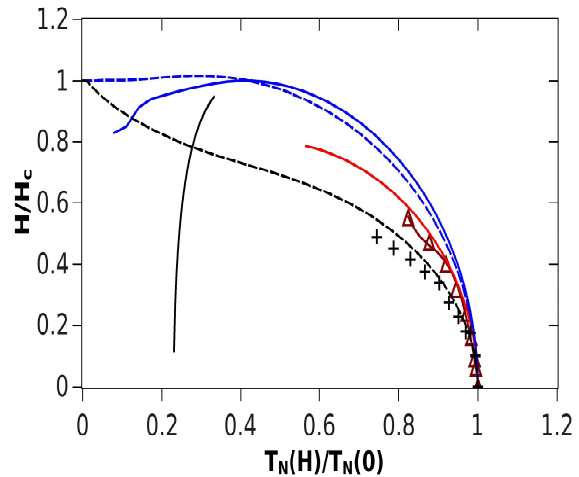}
\caption{The dependence of the reduced magnetic field $H/H_{c}$ as a function N\'eel temperature $T_{N}(H)/H_{N}(0)$ of the quantum isotropic 
spin-$1/2$ Heisenberg antiferromagnetic model on bcc lattice $(z=8)$. The black, red and blue continuos lines are MFRG-12, EFRG-12 and 
EFRG-24 schemes, respectively, are our contributions. The black dashed line represent MFA-2 scheme \cite{bublitzbcc}, the blue dashed line is 
EFT-4 scheme \cite{minos2012}, the marks in cross represents the results of high-temperature series expansions \cite{pan} and the marks in wine 
triangule represents the experimental results \cite{freitas2015}. The wine line represented is a fit third-order polynomial as explained in 
the text.} 
\label{figure5}
\end{figure}

In Fig. (\ref{figure5}) the critical frontier calculated by MFRG-12, EFRG-12 and EFRG-24 for the bcc lattice is compared with that obtained by 
high-temperature series expansions \cite{pan}. Only EFRG-12 and EFRG-24 shows qualitatively the same behavior. The EFRG-24 curve exhibits a 
reentrant zone in the interval  $ 0.04 < T_{N}/T_{N}(0) < 0.82$, when the field is between $ 0.83 < H/H_{c} < 1.0$. We have also applied the 
MFRG-12 approach, but for larger values of $T$ there are no critical fields solution curve that shows a spurious criticality. 

A zero field, EFRG-12 gives $T_{N}(H=0) \simeq 6.31$ (in $J/k_{B}$ units) and EFRG-24 we have $T_{N}(H=0) \simeq 6.29$ (in $J/k_{B}$ units). This numerical results obtained is also compared with other different methods, such as high-temperature expansion $T_{N}(H=0) \simeq 5.53$ \cite{pan}, 
MFA-2 $T_{N}(H=0) \simeq 7.83$ \cite{bublitzbcc}, EFT-2 $T_{N}(H=0) \simeq 6.94$ , and EFT-4 $T_{N}(H=0) \simeq 6.89$ \cite{minos2012}. The 
maximum point only occurs EFRG-24, where the values is $(2.507,6.695)$. This value is non-normalized. In this lattice there is an inflection 
point around $\left(T_{N}/T_{N}(0)\right)\cong 0.118$. 

In order to compare with some experimental results, in the low-field limit, we obtain a critical behavior similar (qualitatively) for the 
$BaCuO_{2}$ compound \cite{genoud1995}, i.e., $T(H)/T_{N}(0)\simeq 1+\tilde{a}H^{2}+\tilde{b}H^{3}$ with $\tilde{a}\simeq-0.59$ mK/T$^{2}$ 
and $\tilde{b}\simeq-0.104$ mK/T$^{3}$ using EFT-4 \cite{minos2012} (while the experimental results are $a\simeq-0.51$ mK/T$^{2}$ and 
$b\simeq-0.11$ mK/T$^{3}$, with $T_{N}(0)=11.97$ K) and using EFRG-24 (present result) the values are $\tilde{a}\simeq-0.57$ mK/T$^{2}$ 
and $\tilde{b}\simeq-0.105$ mK/T$^{3}$. For the low-anisotropy antiferromagnet $Cs_{2}FeCl_{5}\ldotp H_{2}O$ compound \cite{freitas2015} 
the critical behavior at low-field limit is the type $T(H)/T_{N}(0)\simeq 1+\tilde{c}H+\tilde{d}H^{2}+\tilde{e}H^{3}$ with 
$\tilde{c}\simeq-0.430$ mK/T, $\tilde{d}\simeq 0.503$ mK/T$^{2}$ $\tilde{d}\simeq-0.181$ mK/T$^{3}$ using EFRG-12 (present result) 
(while the experimental results are $c\simeq-0.434$ mK/T, $d\simeq 0.491$ mK/T$^{2}$ and $e\simeq 0.185$ mK/T$^{3}$, 
with $T_{N}(0)=6.31$ K).

\section{Conclusions}

In this paper the isotropic Heisenberg antiferromagnet has been studied subjected to a longitudinal magnetic field. Our contribution has consisted 
in  the application of mean-field and effective-field renormalization group techniques for finding the critical frontier in the reduced field 
$T_{N}/T_{N}(0)-H/H_{c}$ plane. 

We implemented  MFRG-12, EFRG-12 and EFRG-24 for the simple and the body-centered cubic lattice. The resulting curves for the sc lattice show that the 
MFRG frontier line has a very sharp reentrance, which is softening when the sizes of the clusters grow in the EFRG technique. It is very likely that 
the exact curve lacks of that reentrance, and  the frontier line touches perpendicularly  the field axis. This last geometric detail can be observed in 
Monte Carlo results for an Ising antiferromagnet in the presence of a magnetic field in the sc lattice \cite{salmon}.

Nevertheless, the three curves agree qualitatively for lower fields (see Fig. \ref{figure4}). In what the bcc lattice concerns, the frontiers obtained by 
EFRG-12 and EFRG-24 are similar to that fitted with the data points calculated by high-temperature series expansions \cite{pan}, only at lower fields (see Fig. \ref{figure5}). However, the EFRG-12 curve has no solution at low temperatures, and only the EFRG-24 frontier ends very close to the field axis with some rentrancy. On the other hand, the MFRG-12 approach did not work for this lattice. 

On the other hand, the EFRG-24 method with increasing of the magnetic field we have a \textit{crossover} from high to low temperature that 
is indicated by the change of the curvature $T_{N}$ (inflection point), i.e., $T_{N}$ concave if $H/H_{c}<1.1$ (for sc lattice) and $H/H_{c}<0.89$ (for bcc lattice). We therefore verify that the limit of low fields in real (experimental) situations proves that the critical behavior is a polynomial of order 3, while the results of MFA-2 \cite{bublitzbcc} and series expansions \cite{pan} show a behavior of the type $T_{N}(H) = T_{N}(0) [1-B(H/{zJ})^{2.08}]$.

Based on a previous study, we believe that the model analyzed in this paper will present a richer critical behavior when implemented system with competing short and long-range interactions in the Heisenberg-Kitaev model \cite{lukas2016}. This type of behavior has already been studied by some theoretical models \cite{minos2015a, minos2015b} in the analysis of organometallic compounds. This opens the possibility for further studies, where we can find experimental data to compare with the theoretical results. 

\textbf{ACKNOWLEDGEMENT}

This work was partially supported by CNPq and FAPEAM  (Brazilian Research Agencies).

\end{document}